Perspective | 🔓 Open Access |

# Electron Correlation: Nature's Weird and Wonderful Chemical Glue

Prof. Dr. Jan M. L. Martin ✉

## Abstract

It can be argued that electron correlation, as a concept, deserves the same prominence in general chemistry as molecular orbital theory. We show how it acts as Nature's "chemical glue" at both the molecular and supramolecular levels. Electron correlation can be presented in a general chemistry course in an at least somewhat intuitive manner. We also propose a simple classification of correlation effects based on their length scales and the size of the orbital gap (relative to the two-electron integrals). In the discussion, we also show how DFT can shed light on wavefunction theory, and conversely. We discuss two types of "honorary valence orbitals", one related to small core-valence gaps, the other to the ability of empty 3d orbitals in 2nd row elements to act as backbonding acceptors. Finally, we show why the pursuit of absolute total energies for their own sake becomes a sterile exercise, and why atomization energies are a more realistic "fix point".

# Electron Correlation: Nature's Weird and Wonderful Chemical Glue

Jan M. L. Martin*[a]

*To the memory of Prof. S. Roy Caplan (1927–2021), scientist and polymath*

**Abstract:** It can be argued that electron correlation, as a concept, deserves the same prominence in general chemistry as molecular orbital theory. We show how it acts as Nature's "chemical glue" at both the molecular and supramolecular levels. Electron correlation can be presented in a general chemistry course in an at least somewhat intuitive manner. We also propose a simple classification of correlation effects based on their length scales and the size of the orbital gap (relative to the two-electron integrals). In the discussion, we also show how DFT can shed light on wavefunction theory, and conversely. We discuss two types of "honorary valence orbitals", one related to small core-valence gaps, the other to the ability of empty 3d orbitals in 2nd row elements to act as backbonding acceptors. Finally, we show why the pursuit of absolute total energies for their own sake becomes a sterile exercise, and why atomization energies are a more realistic "fix point".

## 1. Prologue

*If I were forced to sum up in one sentence what the Copenhagen interpretation [of quantum mechanics] says to me, it would be "Shut up and calculate!"* – N. David Mermin.[1]

Nearly all undergraduate general chemistry classes teach molecular orbital theory at some level – unless they defer it to a physical chemistry or quantum chemistry class. Electron correlation is not generally taught as part of general chemistry courses – yet I believe it ought to be, as its crucial importance for chemical bonding and transformations is not widely enough appreciated. Below I will attempt to sketch a way of doing so, and along the way digress into a few aspects of electron correlation that deserve broader appreciation even within the computational chemistry community.

## 2. A Traffic Metaphor

"All models are wrong, but some are useful", as the great statistician George Box put it. One such useful model is Hartree-Fock theory, which indeed is the origin of the entire concept of molecular orbitals.

Faced with the (at the time) complete intractability of the time-independent Schrödinger equation for more than two electrons, Hartree[2] instead treated a model system in which the electronic motions are statistically independent: the probability of finding two particles in their position is the product of the individual probabilities. Hence the N-particle wave function becomes a simple Hartree product of 1-particle wave functions, which in an atom one could call atomic orbitals, and in a molecule, molecular orbitals.

Allow me to give a real-world metaphor for this model. Imagine that all cars driving in downtown Tel-Aviv would have blackened windows, no headlights, and that their drivers were unable to see any other *individual* vehicle or avoid it in real time – but the drivers would see *traffic density* on a Waze screen and be able to steer away from crowded streets and lanes. Welcome to Hartree World.

Now unlike motor vehicles, electrons don't have license plate numbers; hence the probability density, $|\psi^*\psi|$ should be invariant under a particle permutation P. This latter constraint can only be solved in two possible ways: either (for bosons, integer-spin particles) $P\psi = \psi$, or (for fermions, half-integer spin particles like electrons) $P\psi = -\psi$. (About the origin of this distinction, Feynman famously quipped: 'I couldn't reduce it to the freshman level. That means we don't really understand it.'[3]) Fock[4] and Slater[5,6] simultaneously and independently extended Hartree theory for this latter scenario, and created Hartree-Fock theory, with the Slater determinant as the associated model wave function.

We could extend our Tel-Aviv traffic model in a somewhat contrived fashion to feature all identical cars without license plates, and divided roads with the median corresponding to the division between spin-up and spin-down electrons.

[a] *Prof. Dr. J. M. L. Martin*
Department of Molecular Chemistry and Materials Science, Weizmann Institute of Science, 7610001 Reḥovot, Israel
phone: +972 8 9342533
fax: +972 8 9343029
E-mail: gershom@weizmann.ac.il

In the real world, cars of course do avoid each other instantaneously, either through the agency of their human drivers, or increasingly with AI-based assisted-driving features. In such a world, obviously there will be a nonzero difference $|\psi_{12}^*\psi_{12}| - |\phi_1^*\phi_1||\phi_2^*\phi_2|$ – a correlation hole, if you like, as the movements are now statistically correlated.

## 3. How Important Is Correlation Really?

In this regard, wavefunction theory (WFT) can learn a thing or two from density functional theory (DFT). For neutral spherical atoms with large atomic number Z, an asymptotic series in Z can be derived for the total energy (in atomic units, a.k.a., hartree, $E_h$):[7,8]

$$E_{TF} = -0.7687\, Z^{7/3} + 0.5\, Z^{6/3} - 0.0491\, Z^{5/3} + .... \quad (1)$$

(where TF stands for the Thomas-Fermi approximation[9–12])

$$E_{HF} = -0.7687\, Z^{7/3} + 0.5\, Z^{2} - 0.2699\, Z^{5/3} + .... \quad (2)$$

Surprisingly, as Schwinger[13,14] put it, this latter expression is "unreasonably accurate" across the Periodic Table, with even hydrogen (Z=1) only being 10% in error!

Elliot and Burke derived an expression for the exchange energy component:[15]

$$E_x = -0.2208\, Z^{5/3} - 0.2240\, Z + 0.2467\, Z^{2/3} + .... \quad (3)$$

Of which the leading coefficient corresponds to the local density approximation (LDA) and the linear coefficient to the gradient correction.

A similar, more empirical expression for the correlation energy was published in 2016 by Burke et al.[16]

$$E_{c,'exact'} = -0.02073\, Z\, \ln Z + 0.0378(9)\, Z + .... \quad (4)$$

What can we learn from these expressions? First of all, absolute total energies of atoms, let alone molecules, are humungous in chemical terms, considering that 1 $E_h$ = 627.5095 kcal/mol. The pursuit of exact reproduction of total electronic energies appears to be an exercise in futility to this writer (see also Section 3); in contrast, total atomization energies (TAEs), ionization potentials, electron affinities, and the like are on energy scales that one can at least chemically grasp.

Second, the relative importance of exchange and correlation energies has the leading term:

$$E_x/E_c = 10.66\, Z^{2/3}/\ln Z + .... \quad (5)$$

Third, the ratio between correlation and total energy has the leading term:

$$E_c/E = 0.027\, \ln Z/Z^{4/3} + .... \quad (6)$$

This means, correlation energy will be a fraction of a percent of the total energy. So why do we even care?

Well, 0.3% of a huge number is still a pretty large number: for example, the exact valence correlation energy of neon[17] is $-0.32214\, E_h$, or $-202.1$ kcal/mol. But if this error stayed more or less constant between molecules and separate atoms, we would not lose too much sleep over it, would we?

In Table 1, the reader can find, for a number of common small molecules, the percentages of the total atomization energy (i.e., the sum of all bond energies) accounted for by electron correlation. None of the entries are less than 20%, and a few such as $F_2$ and $O_3$ even exceed 100% – i.e., dissociation of these molecules is exothermic (!) at the Hartree-Fock level, which is plainly absurd. In general, if a single Slater determinant is a good zero-order representation of the wave function, we get closer to the 20% end, and if it becomes a poor zero-order representation (i.e., the molecule exhibits static correlation, *vide infra* Section 4.2), the percentage shoots up. Remember always that quantum chemistry is one big exercise in small(-ish) differences of very large numbers.

**Table 1.** Component breakdown (kcal/mol) of the total atomization energies of five representative molecules from the W4-17 dataset.[40]

| Component | $CH_4$ | $SiF_4$ | $C_6H_6$ | $N_2$ | $O_3$ |
| --- | --- | --- | --- | --- | --- |
| SCF | 331.55 | 448.41 | 1045.34 | 119.69 | −45.09 |
| Of which exchange[c] | 168.5 | 359.75 | 583.3 | −70.1 | −199.2 |
| Valence CCSD | 84.71 | 119.07 | 290.34 | 98.09 | 163.94 |
| Valence (T) | 2.89 | 10.12 | 26.76 | 9.46 | 25.62 |
| CCSDT−CCSD(T) | −0.10 | −1.14 | −2.27 | −0.79 | −1.34 |
| Q, 5, ... | 0.11 | 0.93 | 1.78 | 1.19 | 4.38 |
| Inner-shell corr. | 1.28 | 0.84 | 7.50 | 0.84 | −0.05 |
| Scalar relativistics | −0.19 | −1.90 | −0.99 | −0.14 | −0.25 |
| Spin-orbit coupling | −0.08 | −1.97 | −0.51 | 0.00 | −0.67 |
| DBOC | 0.05 | 0.05 | +0.23 | 0.01 | −0.03 |
| ZPVE[a] | 27.74 | 8.04 | 62.08 | 3.36 | 4.15 |
| Total at 0 K | 392.46 | 565.95 | 1306.07 | 225.00 | 142.40 |
| Experiment (AtcT)[b] | 392.47(2) | – | 1306.10(7) | 224.94(1) | 142.48(1) |

[a] Zero-point vibrational energy (see, e.g., Ref.[41,42]) [b] From http://atct.anl.gov version TN 1.122r [c] Present work.

Now why does electron correlation act as such a strong "chemical glue"? Let's go back to our Tel Aviv traffic model: imagine it is rush-hour and more cars are on the road. Obviously 'instantaneous evasion' behavior (accompanied by signaling, honking, and perhaps some swearing in Arabic, Yiddish, or Russian) will increase as traffic density increases.

Does it work the same way with electrons? The zero-order approximation DFT people rely upon most is the homogenous electron gas. For this so-called local density approximation, compact expressions for the correlation energy density exist:[18–21] the simplest and most elegant is perhaps due to Chachiyo:[21]

$$\varepsilon_c(r_s) = a \ln(1 + b/r_s + b/r_s^2) \quad (7a)$$

$$a = (\ln 2 - 1)/2\pi^2 = -0.015545 \, E_h, \quad b = 20.4562557 \quad (7b)$$

Where $\varepsilon_c(r_s)$ is the correlation energy density and $r_s = (4\pi\rho/3)^{-1/3}$ is the Wigner-Seitz radius, i.e., the radius of a sphere whose volume is given by the density in terms of particles per volume. This means in practice that in the low-density limit, the correlation energy density is proportional to $\rho^{1/3}$, while in the high-density limit, it becomes proportional to $\ln \rho$.

Now obviously, if we bring atoms together into a molecule, there will be a region where the electron density will increase: the bonding region. In Bader's QTAIM (quantum theory of atoms in molecules),[22] the electron density will exhibit a 'bond critical point' along the bond axis where the electron density gradient is zero: it is a minimum along the bond axis, but a maximum in the 'zero-flux surface' the bond axis goes through.

What happens to the correlation energy? By way of illustration, consider a "molecule" consisting of two $1s$ Slater orbital densities $A.\exp(-\zeta|\mathbf{r}-\mathbf{r}_A|)$ centered at xyz coordinates $(0,0,\pm 0.5)$. For each "atom" alone, the correlation energy will be given by

$$E_{c,A} = \int \rho \varepsilon_c(\rho) d\tau \quad (8)$$

And similarly for the other atom, $E_{c,B}$, while in the "promolecular[23,24] approximation" $\rho_{AB} \approx \rho_A + \rho_B$,

Hence:

$$E_{c,AB} = \int \rho_{AB} \, \varepsilon_c(\rho_{AB}) d\tau \quad (9)$$

That is (collecting the integrands under the same integral):

$$\Delta E_{c,AB} = E_{c,AB} - E_{c,A} - E_{c,B} \quad (10a)$$

$$\Delta E_{c,AB} = \int [\rho_{AB} \, \varepsilon_c(\rho_{AB}) - \rho_A \, \varepsilon_c(\rho_A) - \rho_B \, \varepsilon_c(\rho_B)] d\tau \quad (10b)$$

If we plug Eq.(7) into Eq.(10), in the low-density limit the integrand will be a *negative* constant multiplied by $(\rho_A + \rho_B)^{4/3} - \rho_A^{4/3} - \rho_B^{4/3}$, which using the generalized binomial theorem can easily be shown to be positive for all $\rho_A, \rho_B \geq 0$. In the high-density limit, we have a negative constant times $(\rho_A + \rho_B)\ln(\rho_A + \rho_B) - \rho_A \ln \rho_A - \rho_B \ln \rho_B = \rho_A \ln(1 + \rho_A/\rho_B) + \rho_B \ln(1 + \rho_B/\rho_A) \geq 0$. This is illustrated below in Figure 1.

Does this continue to be the case *beyond* the LDA? The simplest and most elegant post-LDA expression of the correlation energy density is due to Becke.[25] For opposite spins

$$\varepsilon_{c,\alpha\beta} = \varepsilon_{c,\alpha\beta,LDA}/(1 + 0.0031(\chi_\alpha^2 + \chi_\beta^2)) \quad (11a)$$

while for parallel spins

$$\varepsilon_{c,\sigma\sigma} = \varepsilon_{c,\sigma\sigma,LDA}/(1 + 0.038\chi_\sigma^2)^2 \, [D/D_{\sigma,UEG}] \quad (11b)$$

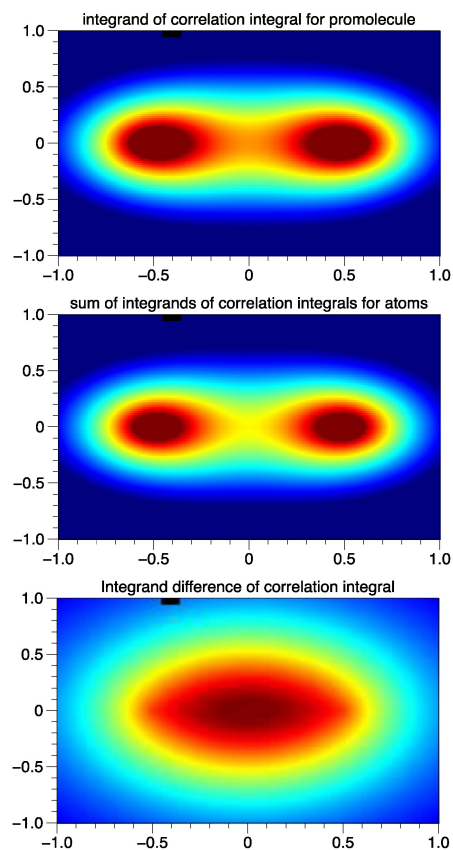

**Figure 1.** Heatmap plot in the xz of the integrand of Eq. (10b) for a promolecule made up of two Slater-type "atomic" densities at $\pm 0.5$ on the z axis.

where the x are reduced density gradients and the D factor in the same-spin expression ensures that a one-electron system has no spurious self-correlation energy.

In other words, the enhancement factor, i.e. the ratio of GGA and LDA correlation energy densities, will be largest where the density gradient is smallest – again, this condition is fulfilled near the bond critical point.

## 4. The Faces of Correlation

### 4.1 Form of The Exact Solution in a Finite Basis Set

According to Löwdin's theorem,[26] the exact wavefunction within a given finite basis set can be expressed as a linear combination of all the possible excited determinants that can be generated from the ground state ("full configuration interaction" or FCI). As the computational cost of FCI scales factorially with the numbers of electrons $n$ and of basis functions $N$, this is not a practical option except for the very smallest systems.

Møller-Plesset many-body perturbation theory (MBPT) is one option. The HF determinant is an eigenfunction of the sum of all one-electron Fock operators $\Sigma_i F_i$. If we consider the difference with the full Hamiltonian to be a perturbation, $V = H - \Sigma_i F_i$, then $E(0) = \Sigma_i \varepsilon_i$, and $E(0) + E(1) = <0|\Sigma_i F_i + V|0> = <0|H|0> = E_{HF}$. In second order (see below):

$$E^{(2)} = \frac{1}{4} \sum_{ij,ab} \frac{\langle ij||ab\rangle^2}{\varepsilon_i + \varepsilon_j - \varepsilon_a - \varepsilon_b} \quad (12)$$

Where $ij$ and $ab$ are occupied and unoccupied ("virtual") orbital indices, respectively, $<ij||ab>$ is a two-electron repulsion integral in the MO basis, and $D_{ijab} = \varepsilon_a + \varepsilon_b - \varepsilon_i - \varepsilon_j$ is a perturbation theory denominator. While MP2 ($2^{nd}$ order MBPT) has only a single term, the number of terms rapidly mounts with increasing orders, and becomes unwieldy at 5th and $6^{th}$ order.[27] Unfortunately (see below), in situations where some of these denominators are very small, convergence of the MP perturbation series is very slow, or the series may diverge.

Limited CI minimizes the energy with respect to the linear coefficients of certain subclasses of determinants (most commonly: CISD or CI with all single and double excitations). Alas, in such a "linear ansatz", the energy is not size-consistent: it can be easily seen, for example,[28] that the exact CISD wavefunction of two He atoms at infinite distance would be (aside from antisymmetrization) the product of two He atom CISD wavefunctions, and hence entail the product of simultaneous and independent double excitations – "disconnected quadruple excitations". Already for two nitrogen atoms, the size-consistency error is on a similar order of magnitude as the bond energy.

The problem can be eliminated completely by enforcing inclusion of all disconnected products – i.e., by using an "exponential ansatz". This is known as coupled cluster (CC) theory, which has emerged as by far the most effective single-reference correlation approach. Originating in nuclear physics, its application to quantum chemistry was first proposed by Čížek and Paldus.[29–31] While the group of Prof. Bartlett has become most identified with the development and practical application of this approach, one must point out the pivotal role of Prof. Isaiah Shavitt z"l (formerly of the Technion) in the early development of coupled cluster theory.[32] Bartlett and Shavitt co-authored a priceless monograph on many-body methods in quantum chemistry.[33]

The CCSD(T) method,[34,35] in which single and double substitutions are treated exactly and triple substitutions approximated quasiperturbatively, is a particularly good compromise between accuracy and computational cost, and has become known as "the gold standard of quantum chemistry" (a term first used by Thom H. Dunning, Jr. in a 2000 lecture). It is effectively a "Pauling point" in that it benefits from mutual cancellation between two neglected terms (Refs.[36–38] and references therein): higher-order triple substitutions (almost universally repulsive) and connected quadruple and higher substitutions (universally attractive). Stanton[39] offers a different perspective why CCSD(T) works.

The CPU time of canonical CCSD(T) scales as $N^7$ with system size. Close approximations that scale only linearly with system size can be obtained, however, through localized natural orbital approaches, such as those of the Neese,[43–45] Werner,[46] and Kállay[47–49] groups.

While the computational cost of CCSDT(Q) and CCSDTQ calculations remains daunting, recent advances[50–53] have made them possible at least for small molecules.

### 4.2 Dynamical and Static Correlation

If in the second-order correlation energy expression the HOMO-LUMO gap is sufficiently large, so will the corresponding denominator in Eq.(12) be, and the n-particle wave function will be dominated by the Hartree-Fock reference determinant. This is said to be dynamical correlation. The correlation energy of the He-like atoms ($H^-$, He, $Li^+$, $Be^{2+}$,…) is an archetypical example: as Z increases, $E_{corr}$ will approach a constant limiting value. Even as a single Slater determinant will of course not recover the correlation energy, it will remain a good *zero-order approximation* to the wavefunction.

At short distance, this type of correlation may be quite important numerically, but a method like CCSD(T) will yield "gold standard" results, while semilocal DFT functionals will handle this regime fairly well. (The higher and more slowly varying the density, the better semilocal DFT works.) As illustrated in Figure 2, the correlated density will be qualitatively similar to the Hartree-Fock density.

Between occupied orbitals separated by a considerable distance (e.g., between monomers in a noncovalent dimer), the two-electron integrals will be small, but many such contributions will add up to a *dispersion energy* that may be strong enough to (help) overcome exchange repulsion. (See Section 4.6 below.) Semilocal correlation functionals intrinsically will struggle to recover this, but even MP2 can generally handle dispersion quite well (owing to an error compensation between neglect of the – usually repulsive – third-order correction, and of the attractive fourth-order triples contribution).

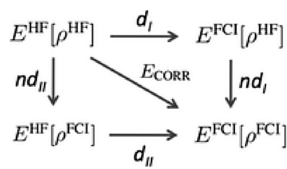

**Figure 2.** Separation between static (nd) and dynamical (d) correlation. CC:BY 3.0 from Ramos-Cordoba et al.[54]

For small HOMO-LUMO gaps (at short or long distance), the corresponding denominators in Eq.(9) will be very small, leading to a near-singularity in the expression and poor convergence of the MBPT expansion. The n-particle wave function, for its part, will contain one or more prominent excited determinants aside from the Hartree-Fock reference. This phenomenon is known as near-degeneracy correlation, a.k.a. static correlation, a.k.a. nondynamical correlation – a rose by any other name smells just as computationally demanding.

The archetypical example of absolute near-degeneracy correlation (a.k.a., "type A static correlation"[55]) is probably stretched $H_2$: as one stretches further and further, the wavefunction will get ever closer to biconfigurational character, and will be purely biconfigurational at long distance. A single determinant, even though it is a good zero-order wave function near equilibrium, will become a qualitatively incorrect zero-order wavefunction (with a qualitatively incorrect density, see Figure 2) as the bond is stretched and the gap narrows.

Another hallmark of type A static correlation is RHF/UHF instability or, in a UHF calculation, "spin contamination" (i.e., the expectation value of $\hat{S}^2$ exceeds that of the pure spin state due to admixture of a determinant with the next higher spin). The onset of such instability as a bond is stretched was first pointed out by Coulson and Fischer,[56] hence the name 'Coulson-Fischer point'. But molecules like singlet BN, singlet $C_2$, and $O_3$ are UHF-unstable even at their equilibrium geometries.

Hollett and Gill[55] distinguish between type A static correlation (absolute near-degeneracy) and type B static correlation (relative near-degeneracy). The latter is best illustrated by HG's paradigmatic example: the Be isoelectronic series $Li^-$, Be, $B^+$, $C^{2+}$, $N^{3+}$,... which has a narrow HOMO-LUMO gap between the 2s and 2p orbitals. As Z increases, the gap actually does not narrow in absolute terms – but it narrows in relative terms, as the wavefunction becomes more and more compact and the squared integrals in the numerator increase steeply. Indeed, in the large Z limit, $E_{corr}$ does not converge to a constant like in the He-like series, but becomes *linear* in Z.

For $Z=3$ (i.e., $Li^-$) you have type A static correlation (because of 2s, 2p near-degeneracy) and a UHF instability, but as Z is varied continuously, a 'Coulson-Fischer-like' point is reached around $Z=4.2$, and beyond that the RHF solution is stable with respect to spin-symmetry breaking, but type B static correlation keeps growing more important. While this is an intriguing "boutique" phenomenon, in "real-life" chemical space type A correlation is a much more widespread problem.

Multireference methods are generally the approach of choice for very strong type A static correlation; however, as we have shown extensively,[37,38] single-reference coupled cluster approaches can still hold their own if the cluster operator is adequately expanded. Fortunately, as we were able to show, the further up in the cluster expansion that you walk, the more rapid basis set convergence for the successive cluster terms becomes.[38] (This actually starts happening already for the (T) triples,[57] which has been exploited to mitigate computational cost in our W$n$ series of composite thermochemical methods.[36,37,58,59]) For instance, the effect of connected quintuple excitations is largely recovered even with unpolarized split-valence basis sets.[38] (For a recent, more detailed discussion of basis set convergence of higher excitation terms, see Karton.[60,61]) This phenomenon can be exploited in composite wavefunction-based thermochemistry approaches, as it is in our own W4 theory,[37,38] in the HEAT approach of Stanton and coworkers,[62–65] and in the FPD approach.[66,67]

While a full discussion of basis set convergence would require a separate paper, suffice to say that unlike type A static correlation, the dynamical correlation energy converges *very* slowly with the basis set, roughly as $\propto N^{-1}$;[68,69] if basis sets are built up of complete shells, such as the correlation consistent[70,71] cc-pV$k$Z family, then the number of basis functions in them grows cubically with the cardinal number $k$, and one recovers the familiar $\propto k^{-3}$ inverse-cubic formula,[72] originally rationalized by the partial-wave expansion of atomic pair correlation energies.[73–77] The different convergence behaviors can be intuitively rationalized as stemming from type A static correlation being dominated by relatively few low-lying orbitals, while dynamical correlation results from lots of very small contributions.

A number of diagnostics for static correlation have been reviewed in Ref.[78] These include but are not limited to the largest single and double excitation amplitudes; the vector and matrix norms ($T_1$ and $D_1$, respectively[79,80]) of the single excitations; the slope of the atomization energy with respect to the percentage of exact exchange in DFT;[78] various diagnostics based on natural orbital occupation numbers;[54,81,82] and more. In a very recent study,[83] we have shown that statistically all these diagnostics cluster into three groups: single excitation diagnostics; bandgap diagnostics (themselves weakly subdividing into an entropy group and a double excitations group); and pragmatic energy-based diagnostics. Among the latter, the percentage of the TAE energy accounted for by triple excitations, and the percentage accounted for by correlation energy overall, were earlier shown[37] to be good predictors for the importance of post-CCSD(T) correlation effects.

### 4.3 Inner-Shell and Valence Correlation

It is generally taken for granted that the valence electrons contribute most to chemical properties. According to Nesbet's theorem,[84] the correlation energy can be generally partitioned into (occupied orbital) pair contributions, at least in the closed-shell case. We can then partition the correlation energy into three groups of pairs, according to whether the occupied orbitals are both valence (VV), both inner-shell ("core-core", CC), or one of each ("core-valence", CV).

For second-row and especially heavier elements, the sum of CC and CV energies will actually rival or exceed the valence correlation energy: in practice however, as long as the valence-inner shell energy gap is large enough, the CC energy will mostly cancel between molecule and constituent atoms, while any small chemical effects of inner-shell correlation are largely due to the CV component (at least in the first row: things get a little hairier when you have two adjacent second-row atoms, with their bulkier cores[85]).

As the inner-shell orbitals are typically very compact, the CC and CV energies are comparatively easy to recover, provided the basis set adequately covers the relevant exponent ranges. (For a detailed discussion and analysis of the thermochemical contributions of inner-shell correlation, as well as of their basis set convergence, see Ref.[85] and references therein.)

Note that their impact is also felt in geometrical derivative properties: if one is comparing with geometries and/or vibrational frequencies from high-resolution spectroscopy, core-valence corrections are essential.[86–88]

### 4.4 Honorary Valence Orbitals of the First Kind

A table of the numerical HF orbital energies of the atoms can be found on the author's website[1] and is summarized in graphical form in Figure 3.

It is noteworthy that for alkali and alkali earth metals, at least further down the PTE, the (n-1)s and especially (n-1)p are energetically very close to the valence shell, and indeed those of K and Ca lie *above* the np valence orbitals of chalcogens and halogens (particularly O and F, respectively).

Hence, if one blindly freezes core orbitals, this can cause very large errors in reaction energies involving these atoms, as was noted by Radom[89] and by the present author.[90] The same happens, to a lesser extent, for the (n-1)d orbitals in the heavy p-block metalloids: Bauschlicher[91,92] noted this type of "inversion" for gallium and indium fluorides.

In practice, thus, such orbitals need to be treated as what I would term 'honorary valence orbitals'. (A reviewer pointed out Sections 4.3 and 4.4 closely parallel the analysis in Ref.[93], where even the similar term 'extended valence orbitals' is used. The ORCA program system[94] by defaults correlates these, and only freezes what it calls 'chemical core electrons'.)

In this context, I would like to point out a recent paper by Madsen and coworkers[96] that ascribes the unexpectedly poor

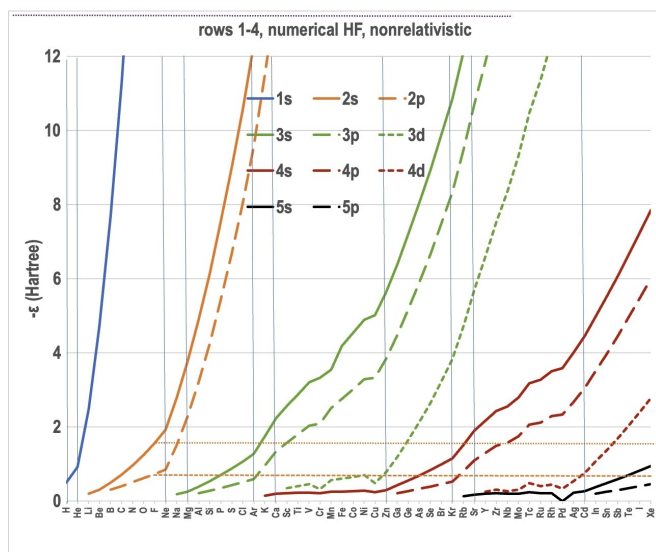

**Figure 3.** Plot of the numerical HF occupied orbital energies[95] of elements H through Xe. Vertical lines delineate s,p,d blocks; the horizontal dotted lines indicate the 2s and 2p valence orbitals of fluorine. Separation between static (nd) and dynamical (d) correlation. CC:BY 3.0 from Ramos-Cordoba et al.[54]

performance of SCAN for alkali metals to a 'semicore' region that is not well captured.

### 4.5 Honorary Valence Orbitals of the Second Kind

While unrelated to correlation as such, the author would like to highlight a second type of honorary valence orbital, as discussed in some detail previously in Ref.[97]

Many years ago, Bauschlicher and Partridge (BP)[98] reported that the B3LYP calculated atomization energy of sulfur dioxide is surprisingly sensitive to the presence of high-exponent *d* functions in the basis set. Later the present author showed[99] that this is in fact the case for other molecular properties as well, including a blue shift on the order of 50 cm$^{-1}$ in the vibrational frequencies.

While BP ascribed the issue to core polarization (see also Ref.[100]), Dunning et al.[101] attributed it to a combination of core polarization and of correlation from the inner loops of the sulfur valence orbitals. However, we were able to show[99,102] that it persists if the core electrons are replaced by an effective core potential (ECP), and indeed that essentially all of the effect is also seen at the SCF level. Hence neither proposed explanation holds water.

The idea that hypervalence was somehow involved (particularly in compounds like $SF_6$ and $PF_5$) has been convincingly refuted by Cioslowski[103] and by Schleyer.[104] (See also Miliordos and coworkers.[105])

---

[1] See https://www.compchem.me/aoenergytable

We did find, over the course of multiple high-accuracy ab initio studies on second-row compounds (e.g.[97,99,102,106,107]), that the degree of sensitivity increases with increasing oxidation state of the central 2nd row atom. In the most extreme such case, namely $Cl_2O_7$,[97] adding tight d functions to an aug-cc-pVDZ basis set will add about 100 (!) kcal/mol to the computed atomization energy, and for $HClO_4$ "only" 50 kcal/mol.

All becomes clear when comparing natural population (NPA) and natural bond orbital (NBO)[108] analyses on $ClO_4^-$ between HF/aug-cc-pVDZ and HF/aug-cc-pVDZ+4d, where "+4d" indicates adding four extra d functions to chlorine with exponents in an even-tempered sequence $\alpha_{d,max} 2.5^n$. The atomization energy increases by 52.9 kcal/mol (!), while in the NBO natural electron configuration, $3d$ occupation on Cl increases from 0.24 to 0.33. A group of five ($3d$)-like NBOs see their aggregate $d$ population increase from 0.28 to 0.37; the five NBOs are split into a quasi-$t_{2g}$ triad with a cumulative population of 0.32 (including some non-d components) and a quasi-$e_g$ dyad with about 0.12. NBO perturbation theory also shows strong interactions between lone pairs on O and the Cl($3d$). In plain English: the 3d orbital on the central 2nd-row atom acts as a backbonding acceptor for chalcogen and halogen lone pairs – the extra d functions are not so much acting as polarization functions as describing the inner part of a valence orbital!

Schaefer and coworkers[109] pointed out a related phenomenon in CaO (see also Ref.[90]).

### 4.6 Dispersion and Intermolecular Interactions

Consider the interactions between two noble gas atoms. At very long distance, the zero-order wavefunction will essentially be an antisymmetrized product of the monomer wave functions. This is also true at the Hartree-Fock level; as we push them closer together, however, we will increasingly see exchange repulsion – this has a very steep distance dependence, modeled as $R^{-12}$ in the Lennard-Jones potential.

But what about the correlation energy? As noted above, we can decompose it exactly into pair correlation terms (trivially for MP2 and CCSD, somewhat less intuitively for higher-order methods[84]). Figure 4 graphically represents the possible scenarios: in practice charge transfer (d–f) will be irrelevant for our example, while (a) will be present in dimer and separate monomers alike. As the atoms come together, (b) will become increasingly important (with the familiar $R^{-6}$ distance dependence from the Lennard-Jones potential) – this is the dispersion term, somewhat mitigated by the exchange-dispersion term (c). The van der Waals distance is the equilibrium point between exchange repulsion and dispersion [attraction].

For more general noncovalent interactions, electrostatic and induction terms are at work, both of which already show up at the Hartree-Fock level, with higher-order corrections from electron correlation.

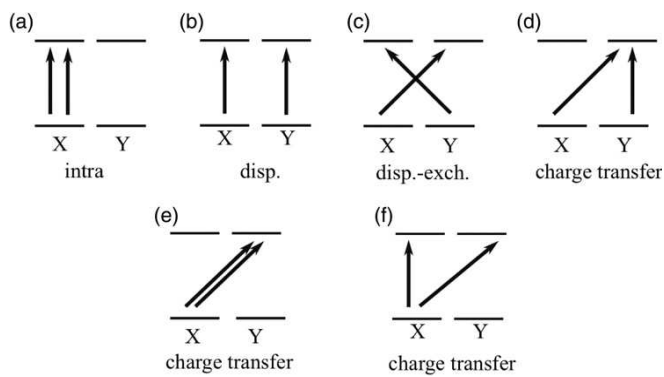

**Figure 4.** Double excitations in an X...Y dimer graphically decomposed into intramolecular, charge-transfer, dispersion, and exchange-dispersion components. CC:BY 4.0 from Bistoni.[110]

A systematic way of studying this is known as SAPT or symmetry-adapted perturbation theory.[111] Grossly oversimplifying, it is a form of perturbation theory in which the zero-order Hamiltonian contains only intramolecular terms, and all intermolecular terms are diverted to the perturbation.

Note that in some rare scenarios (e.g., two water molecules at long distance, where correlation reduces the dipole moments and hence the electrostatic attraction) the *overall* correlation contribution to the noncovalent interaction energy *can* become negative (Ref.[112] and references therein.)

### 4.7 An Attempted Simple Classification of Correlation Effects

An admittedly simplistic "classification quadrant" for correlation effects can be drawn up as follows (Table 2). On one axis, we have the size of the band gap relative to the two-electron integral; on the other, whether the correlation is short-range or long-range.

Dynamical correlation at short range is relatively easy to capture, by either a semilocal DFT correlation functional or by many-body perturbation theory.

Dynamical correlation at long range is intrinsically poorly described by semilocal functionals, yet even MP2 can capture it quite well. Its chemically most important manifestation is dispersion.

**Table 2.** Simplistic classification quadrant for correlation types.

|  | Short-range | Long-range |
| --- | --- | --- |
| Large relative gap | **Dynamical correlation** DFT and MBPT can handle well | **Dispersion** (interatomic) Poorly handled by DFT; low-order MBPT can handle well |
| Small relative gap | **Type B static correlation** Coupled cluster theory; not captured by DFT | **Type A static correlation** Multireference methods; high-order coupled cluster theory DFT (partly) |

Type A static correlation, or absolute near-degeneracy, requires either multireference methods or fairly high-order coupled cluster theory. To some degree, DFT exchange can cope with it: as Cohen and Handy[113] (who point to the earlier work of Baerends[114]) have shown, a strict separation between DFT exchange and type A static correlation is not possible.

Finally, consider the paradigmatic example of type B static correlation arises in the Be isoelectronic series,[55] where the (2s)−(2p) gap may keep growing in absolute terms, but DFT exchange will not recover any type B static correlation in this system, as can be readily verified.

### 4.8 Double-Hybrid DFT Functionals in This Perspective

In the early 1990s, Görling and Levy[115] proposed a perturbation theory expansion in a basis of Kohn-Sham orbitals. (We should stress here that, while it is *analogous* to the commonly used Møller-Plesset perturbation theory in a basis of HF orbitals, GLPT2 is by no means *equivalent* to MP2.)

Double hybrid DFT (see Ref.[116] in this journal for a recent review) is really based on a "marriage of convenience" between the felicitous properties of semilocal DFT for short-range correlation, and of MP2 for long-range dynamical correlation. Zhao and Truhlar[117] may have been the first to attempt to do so, by global linear combination of correlation energies from separate MP2 and DFT calculations; the late János Ángyán[118] considered combining the two using range separation, with MP2 at long distance and DFT at short distance. Grimme[119] was the first to turn Görling-Levy perturbation theory into a practical approach; since then, the term 'double hybrid' has effectively become synonymous with this approach. We showed later[120–122] that the accuracy of DH methods can be greatly enhanced by admitting spin-component scaling and an empirical long-range dispersion correction:[123] it also became clear that the semilocal correlation functionals that work best are those that are short-ranged.

One way to look at it, then, would be that at long range one has a mix of GLPT2 correlation and the empirical dispersion correction; at short range, the latter is shut off by its damping function, leaving this range to a mix of semilocal and GLPT2 correlation, while at intermediate range, one has a mix of all three but predominantly GLPT2.

### 4.9 Exchange Energy in This Picture

Of the Hartree-Fock terms, the nuclear repulsion is classical, while the kinetic energy, electron-nuclear attraction, and two-electron Coulomb repulsion all have classical equivalents. The exchange energy does not: it is a purely quantum-mechanical consequence of the Pauli exclusion principle and of the fact that, unlike soccer or basketball players, electrons "don't wear numbers on their backs", as one of my undergraduate professors whimsically put it.[2] And just like one has a correlation hole, one has an exchange hole.[124]

One might consider this a form of correlation as well, although Löwdin's definition of the correlation energy, $E_{corr} = E_{FCI} − E_{HF}$, excludes it by definition. But leaving that aside, what is the impact of the exchange energy on molecular total atomization energies?

As can be seen in Table 1, for molecules like $CH_4$ or $SiF_4$ that are well described by a single reference determinant, exchange is strongly binding and in fact can contribute over 50% of the Hartree-Fock component of TAE. For molecules with strong static correlation, however, like $O_3$, exchange may actually be strongly antibonding. Also for such molecules, Cohen and Handy[113] found that DFT exchange is considerably more binding than Hartree-Fock exchange, and indeed state "we deduce we cannot separate exchange and non-dynamic correlation".[113]

We have exploited this behavior for a diagnostic for static correlation character based on the difference between Hartree-Fock exchange and semilocal exchange from a HF density.[83]

Interestingly, while exchange in the SAPT picture is repulsive, things are different in a basis of dimer Hartree-Fock orbitals. As seen in Figure 5 for the argon dimer, the exchange component then actually becomes more binding as the atoms are pushed together (cf. the way it is for total atomization energies) but this is outweighed by much stronger Coulomb repulsion.

## 5. The Futility of Pursuing Absolute Energies

A surprisingly widespread practice in DFT functional development and validation (e.g., Ref.[125]) is trying to reproduce exact absolute atomic energies. (The latter can be obtained experimentally from summing the ionization potentials and subtracting the relativistic corrections.[126])

While this may speak to the chemist's imagination as a computational target, it suffers from a fundamental problem. Consider once again the asymptotic expansion in the atomic number Z of the atomic energy. If we combine what we have summarized above[7,8,15,16] about the HF and correlation components, we have:

$$E_{total,NR} = −0.7687\ Z^{7/3} + 0.5\ Z^2 − 0.2208\ Z^{5/3} \\ −0.2618(9)\ Z − 0.02073\ Z\ \ln Z + 0.2467\ Z^{2/3} + .... \quad (10)$$

Where "NR" stands for "nonrelativistic". The Z-scaling for the leading scalar relativistic correction has been known since the late 1920s from the solutions of the Dirac equation for hydrogen-like atoms (these were found simultaneously and independently by Charles Darwin's grandson, Charles Galton Darwin[127] and by Walter Gordon[128]):

---

[2] "Elektronen dragen geen rugnummer." (Lucien Van Poucke, 1937–2005)

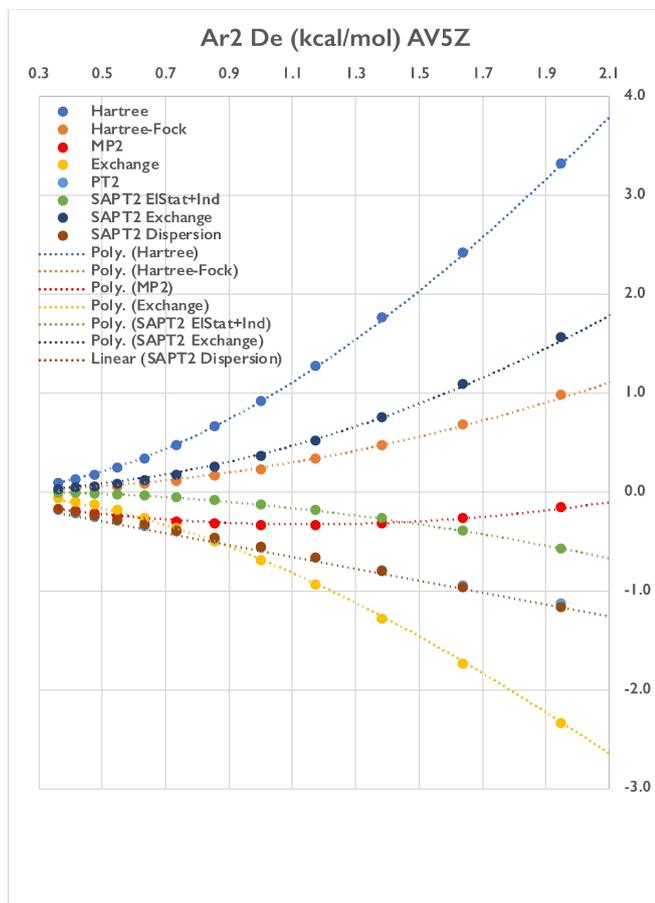

**Figure 5.** Component breakdown of the MP2 and SAPT interaction energies of Ar$_2$ as a function of $(3.8 \text{ Å}/R)^6$.

$$E = mc^2 / \sqrt{1 + \left\{\frac{Z/c}{n - j - \frac{1}{2} + \sqrt{(j+1/2)^2 - \frac{Z^2}{c^2}}}\right\}^2} \quad (13)$$

In which $n$ is the principal quantum number, $m$ the rest mass of the electron, $c \approx 137$ a.u. is the speed of light in atomic units, and the new quantum number $j$ is constructed from the angular quantum number $l$ and the spin quantum number $s$ as $j = \min(l \pm s, 1/2)$.

Expanding $(1+x)^{-1/2}$ as a power series in $x$, one obtains:

$$E = mc^2 - \frac{Z^2}{2n^2} + \frac{Z^4}{2n^4 c^2}\left\{\frac{3}{4} - \frac{n}{j+\frac{1}{2}}\right\} + O\left(\frac{Z^6}{c^4}\right) \quad (14)$$

In which the first term is the Einsteinian rest mass-energy, the second term the nonrelativistic energy of the hydrogen-like atom, and the next term can be further decomposed into a mass-velocity term, a Darwin term, and a spin-orbit term.

Bottom line for our purposes: the relativistic corrections scale as $Z^4$, or if you want to make things clearer, $Z^2(Z/c)^2$. (For larger values of $Z/c$, additional corrections will be necessary.)

Relativistic effects on molecular properties are beyond the scope of this article: for recent reviews, see, e.g., Pyykkö[129] and Schwerdtfeger et al.[130] We do mention in passing that for molecules of lighter elements, scalar relativistic contributions to reaction energies can be surprisingly well modeled in terms of s-orbital, and to a lesser extent p-orbital, populations in an NPA analysis, as we showed in Ref.[131] (at the instigation of Dr. Kenneth G. Dyall). On a personal note, what finally cured me of my skepticism that scalar relativistic effects could be of chemical importance in the first row was a benchmark study of atomic electron affinities,[132] where we found a 0.01 eV contribution already for fluorine (an order of magnitude larger than the difference with experiment).

Coming back to the main issue at hand, it becomes clear that absolute energies quickly take on values many orders of magnitude larger than any reaction energy of chemical interest.

Does that mean that quantum chemistry is doomed? Of course not. Only a tiny morsel of these astronomical energies does not cancel from reactants to products in a chemical reaction – and it is our challenge to single out those bits. But the pursuit of exact absolute atomic energies is best described as a futile, Sisyphean exercise.

So, what is the "next best" if one needs a universal set of thermochemical parameters? The old, established alternative of heats of formation is often associated with experimental conditions that may be difficult to model on the computer. By far the most advanced approach here is that of the Active Thermochemical Tables (ATcT) initiated by Branko Ruscic[133,134] at Argonne National Laboratory: this, however, blurs the line between experimental and theoretical values, as the values are extracted by solving a thermochemical network including all available experimental and theoretical reaction energies in the network (with weights based on their uncertainties[135]).

Researchers in computational thermochemistry often avail themselves of gas-phase total atomization energies, i.e., they take the neutral ground-state atomic energies as their "reference frame": examples include, but are not limited to, the HEAT initiative[64,65] and our own W4-11 and W4-17 datasets.[40,136]

## 6. Summary and Outlook

It can be argued that electron correlation, as a concept, deserves the same prominence in general chemistry as molecular orbital theory. (A similar claim can be advanced on behalf of conceptual DFT.[137]) It is Nature's "chemical glue" at both the molecular and supramolecular levels. And indeed, it can be presented in a general chemistry course without resorting to (too much) abstruse mathematics and physics.

We also propose a simple classification of correlation effects based on their length scales and the size of the orbital gap (relative to the two-electron integrals).

An underlying theme of this essay is that wavefunction ab initio and DFT approaches can learn a lot from each other.

In experimental chemical thermodynamics, the pursuit of absolute enthalpies is eschewed in favor of standard heats of formation. While in principle, one could pursue absolute energies in a computational context, these find themselves so many orders of magnitude larger than any meaningful chemical energy difference that their pursuit for their own sake becomes a sterile exercise. If one needs some sort of anchoring point, total atomization energies (cognates of the experimental heats of formation) are much to be preferred.

## Acknowledgements

This research was supported by the Israel Science Foundation (grant 1969/20) and by the Minerva Foundation (grant 2020/05), Munich, Germany. The author would like to thank (in alphabetical order of first name) Profs. Amir Karton, Branko Ruscic, A. Daniel Boese, Frank Neese, George C. Schatz, Jean Demaison, John F. Stanton, Jürgen Gauss, Kenneth G. Dyall, Kirk A. Peterson, Lars Goerigk, Leo Radom, Martin Head-Gordon, Pekka Pyykkö, Peter Schwerdtfeger, Peter R. Taylor, Sebastian Kozuch, Stefan Grimme, Timothy J. Lee, and Trygve Helgaker for helpful discussions over the years. I would also like to thank Golokesh Santra, Emmanouil Semidalas, and Dr. Nisha Mehta for critically reading the draft manuscript.

Part of this presentation is based on the author's eponymous 2017 Israel Chemical Society Prize lecture.

# PERSPECTIVE

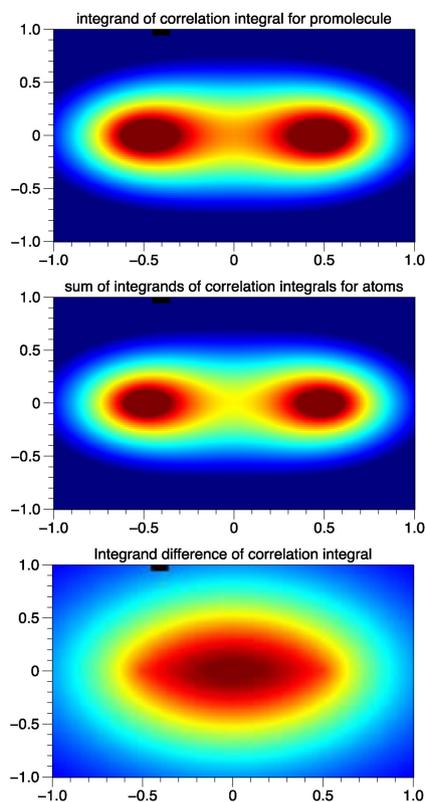

*Prof. Dr. J. M. L. Martin\**

1 – 13

**Electron Correlation: Nature's Weird and Wonderful Chemical Glue**